\documentstyle[12pt,epsf]{article}

\let\section=\subsection     \let\subsection=\subsubsection                

\begin{document}

\begin{center}

   {\large \bf NEUTRINOS FROM PROTONEUTRON STARS}\\[2mm]

     S. REDDY$^a$, J. PONS$^{a,b}$, M. PRAKASH$^a$ AND J. M. LATTIMER$^a$ \\[5mm]
   {\small \it $^a$Department of Physics \& Astronomy, \\
   SUNY at Stony Brook, Stony Brook, NY 11794.\\[8mm] }

   {\small \it $^b$Departament d'Astronomia, Universitat de Val\'encia\\
    E-46100 Burjassot, Val\'encia, Spain.\\[8mm] }
\end{center}
\vspace{-0.1in}
\begin{abstract}\noindent
We study the diffusive transport of neutrinos in a newly born neutron
star to explore its sensitivity to dense matter 
properties.  Energy and lepton number which are trapped during the
catastrophic implosion diffuse out on the time scale of a few
tens of seconds. Results for different dense matter models 
are presented.
\end{abstract}
\vspace{-0.2in}
\section{Introduction}
\vspace{-0.1in} The core of a massive star implodes when its mass
exceeds the Chandrashekar mass. The hot and dense remnant formed
subsequent to the implosion is a protoneutron star. Numerical
simulations of the implosion (and  the subsequent formation of a shock
wave at core bounce) indicate that, due to the high densities and
temperatures, most of the star's gravitational  binding energy and
lepton number released remains trapped in the star as neutrinos. The
general features of the early evolution  have been discussed in prior
work \cite{BL,KJ}. The object of this work is to elucidate the role
played by the microphysical inputs (equation of state (EOS) and
neutrino  opacities) on the macrophysical evolution of the
protoneutron star.
\vspace{-0.2in}
\section{The Basic Equations}
\vspace{-0.1in}
The structure of the star is assumed to be in quasi-static equilibrium, and
is described by the general relativistic equation for hydrostatic equilibrium.
The loss of lepton number and energy due to neutrino flows is treated 
in the diffusion approximation. The equations governing the early evolution
may be found in \cite{BL,KJ}. To solve the structure equations we need to 
specify the finite temperature EOS discussed in \S3. Other microphysical 
inputs required to solve the transport equations are the neutrino mean free 
paths and bulk properties such as the specific heat and the nuclear 
symmetry energy. 
\vspace{-0.2in}
\subsection{Deleptonization}
\vspace{-0.1in}
To  illustrate how microphysical inputs 
influence lepton  number transport, we examine the Newtonian
diffusion equation \cite{RPL} 
\begin{eqnarray}
n\frac{\partial Y_L}{\partial
t}\simeq\frac{1}{r^2}\frac{\partial}{\partial r}
\left[r^2\left((D_2+D_{\bar 2})\frac{\partial(\mu_\nu/T)}{\partial r}
- (D_3-D_{\bar 3})\frac{\partial (1/T)}{\partial r}\right)\right].
\end{eqnarray}
Here $Y_L=Y_e+Y_{\nu}$ is the total lepton fraction and $\mu_{\nu}$ is
the  local neutrino chemical potential. The neutrino mean free path
$\lambda(E_{\nu})$ enters  Eq. (1) via the
unnormalized  diffusion coefficients $D_n$ (and $D_{\bar n}$ for
anti-neutrinos) defined by $D_n =
\int_0^{\infty}dE_{\nu}~E_{\nu}^n~\lambda_\nu(E_{\nu})
~f_\nu(E_{\nu})(1-f_\nu(E_{\nu}))$.  The time scales crucially depend
on the magnitude of $\lambda(E_{\nu})$ and its energy dependence. For
electron neutrinos, the dominant  contributions to $\lambda(E_\nu)$ 
 arise from the charged current  reaction $\nu+n\rightarrow
e^-+p$. The rate of this reaction is sensitive to both temperature
and the  composition of the ambient matter [3-6]. 
Under degenerate conditions, the phase space for this reaction  is
proportional to $\mu_nT^2(\hat{\mu}+E_{\nu})$, where $\mu_n$ is the
neutron  chemical potential and $\hat{\mu}=\mu_n-\mu_p$. The  left
hand side of Eq. (1) illustrates one way the EOS directly influences
the  lepton number flows. Note that  $\frac{\partial Y_L}{\partial
t}=\frac{\partial Y_L}{\partial Y_{\nu}}\frac{\partial
Y_{\nu}}{\partial t}$ , and that $\frac{\partial Y_L}{\partial
Y_{\nu}}$ is directly related to the nuclear  symmetry energy $S(n_B)$. 
Since $\hat{\mu}=4 S(n_b)(1-2Y_P)$, where $Y_p$ is the proton fraction,
 the nuclear symmetry energy will play a
crucial role in the deleptonization phase. Further, the possible
presence of hyperons at high density affects both $\frac{\partial
Y_L}{\partial Y_{\nu}}$ and the neutrino mean free path  $\lambda$
will be altered \cite{RPL,RP}.

\vspace{-0.1in}
\subsection{Cooling}
\vspace{-0.1in}
At early times, the flow of electron neutrinos due to chemical
potential  gradients results in net heating rather than cooling. This Joule
heating and negative temperature gradients warm the central regions
until the center reaches the maximum temperature. At this stage, the
center begins  to cool  as the thermal gradients force all six
neutrino species radially  outwards. Due to  the high densities and 
temperatures present, the pair production processes are sufficiently rapid to 
ensure that $\mu$ and $\tau$ neutrinos are in 
thermal equilibrium. Energy transport is governed by
\begin{eqnarray} 
nT{\partial s\over\partial t}={c\over6\pi(\hbar c)^3}{1\over
r^2}{\partial \over\partial r}\left[r^2\sum_\ell{D_{4,\ell}\over T^2} {\partial
T\over\partial r}\right]. 
\end{eqnarray}
All general relativistic corrections and terms arising due to 
lepton number gradients have been dropped to highlight the essential role
of the microphysics.  The dominant 
opacity for $\mu$ and $\tau$ pairs is that due to  
scaterring on 
baryons and electrons, as the charged current reactions are 
kinematically inaccessible. The charged current reactions are still the 
dominant source of opacity for electron neutrinos; thus, relative to the $\mu$
and $\tau$ pairs their contribution to energy transport is
significantly suppressed.  Under partially degenerate conditions, 
the phase space for scattering reactions  is roughly
proportional to $M^{*2}T^3$ (note that the ambient temperature 
determines the average $E_\nu$);  
$M^*$ arises since it determines the density of states at the Fermi
surface. The left hand side of Eq. (2) is directly related to the 
specific heat since $Tds=C_vdT$, and is strongly dependent on the effective
baryon mass $M^*$. Neutrino mean free paths depend on the same physics, 
since they are proportional to the number of target particles which do not
suffer significant Pauli blocking.
\vspace{-0.2in}
\section{Microphysics}
\vspace{-0.1in}
\subsection{The Equation of State (EOS)}
\vspace{-0.1in}
The dense matter EOS is an important ingredient in
protoneutron star simulations. For a detailed discussion of a variety
of dense matter models see \cite{PR}. In this work, we have explored four 
different models (see Table) to study the early evolution.
We broadly classify them as soft (GM3) and stiff (GM1) models \cite{GM}, 
and models
can contain (GM1-H, GM3-H) or not contain (GM1-N, GM3-N) hyperons. Hyperons 
begin to appear only towards the late stages of the deleptonization \cite{PR}.
This results in these models having significantly lower 
maximum masses for $Y_{\nu=0}$ ($4^{th}$ column) and the 
existence of a range ($3^{rd}$ \& $4^{th}$ columns) of initial masses that 
become unstable upon deleptonization. 
\begin{center}
\begin{tabular}{lcccc}
\hline 
{Model} & Composition & $M_{Y_{L}=0.4}^{MAX}(T=0)$ &$M_{Y_{\nu}=0}^{MAX}(T=0)$   \\
\hline 
GM1-N & n,p + leptons & $ 2.52 $ & $ 2.76 $   \\
GM1-H & n,p,H + leptons & $2.25$ & $1.96$   \\
GM3-N & n,p + leptons & $2.16$ & $2.35$   \\
GM3-H & n,p,H + leptons & $1.95$& $1.74$   \\
\hline
\end{tabular}
\end{center}
\vspace{-0.2in}
\subsection{Neutrino Interactions}
\vspace{-0.1in}
Neutrino interactions are significantly modified in a hot and dense
medium.  The effects of Pauli blocking and strong interactions on the
weak interaction rates have been investigated in Refs. [3-6] 
which showed that density and  temperature dependence of the
neutrino mean free  path are significantly different from  those
employed in earlier proto-neutron star simulations
\cite{BL,KJ}. Strong interaction  effects may be incorporated by
employing  appropriate dispersion
relations and effects of Pauli blocking may be incorporated exactly.  In
addition, Ref. \cite{RPL} shows that the neutrino mean
free path also depends on the composition of the ambient matter.  In
particular, the appearance of hyperons leads to significant reduction
in  $\lambda_{\nu}$ during the cooling phase.
\vspace{-0.1in}
\section{Results}
\vspace{-0.2in}
\begin{figure}[h]
\vspace{-0.1in}
\begin{center}
\epsfxsize=5.5in 
\epsfysize=3.25in
\epsffile{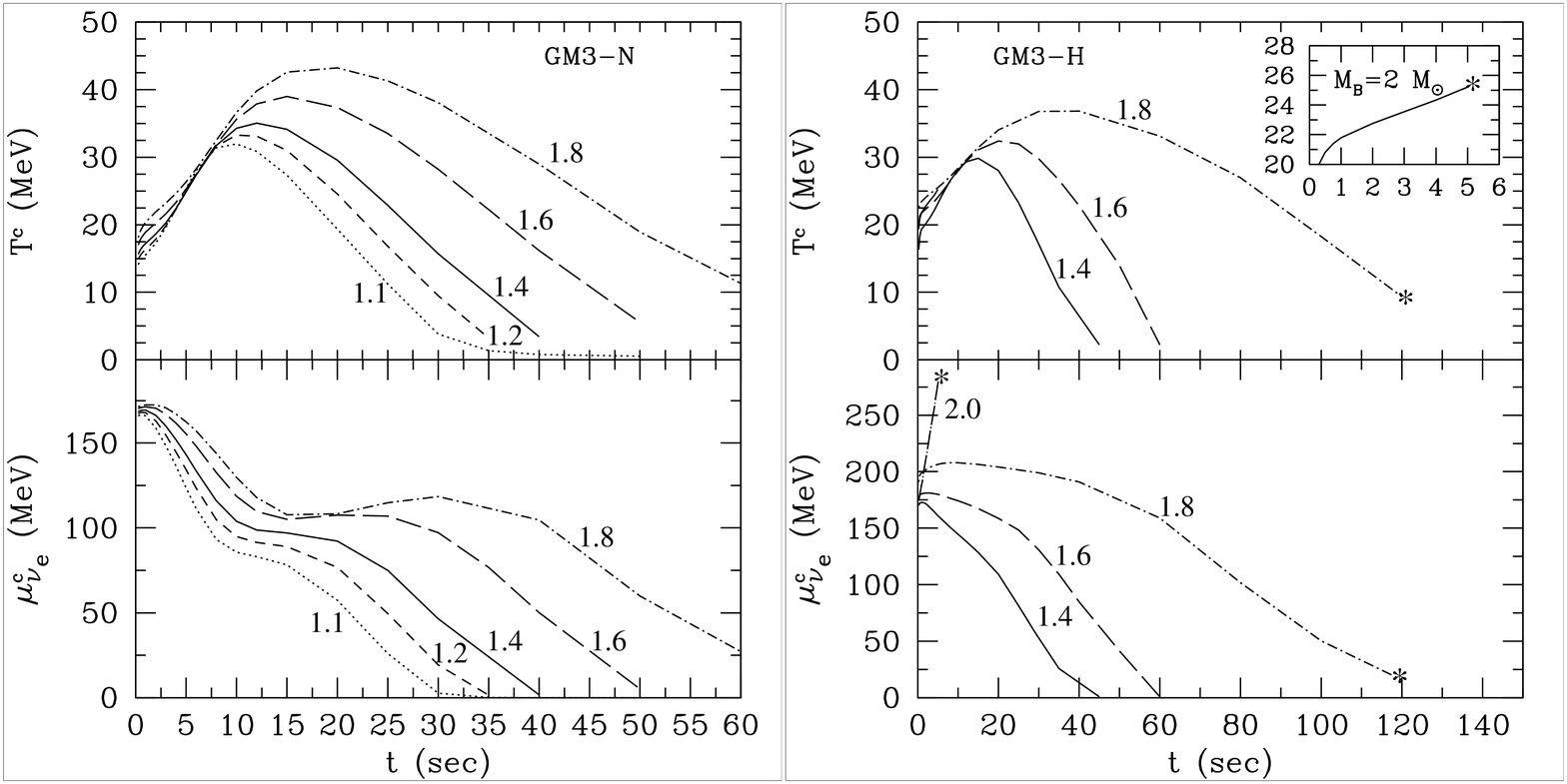}
\caption[]{\footnotesize Central temperature and 
neutrino chemical potential vs time (model GM3). Right (left) panels: 
matter with (without) hyperons. Curves are labelled by the initial
baryon masses. Models $M_B \ge M^{MAX}_{Y_{\nu}=0}=1.74M_{\odot}$ 
are unstable and collapse as they deleptonize (denoted by an asterix). Note, 
however, that if the baryon mass does not significantly exceed this value 
the meta-stable state lasts for a very long time.} 
\end{center}
{\label{center}}
\vspace{-0.1in}
\end{figure}
\vspace{-0.3in}
We turn now to results of simulations which include the full effects
of general relativity. In Fig. 1, the time  evolution of the central 
temperatures are shown for models with (right top) and without (left top) 
hyperons
for different initial masses. The  deleptonization times are related to
the time evolution of the neutrino  chemical potential, which are shown for
models with (right bottom) and without  (left bottom) hyperons.  The 
identification of time scales associated with
deleptonization and cooling depend on their precise definition. Here,
we opt to define the  deleptonization time $\tau_D$ as the time it
takes for the central value of $\mu_{\nu_e}$ to
drop below $10$ MeV, and the cooling time  $\tau_C$ as the time it
takes for the central temperature $T^c$ to drop below  $5$ MeV.
 Fig. 2 
clearly shows the generic trends; (1) Softening leads to higher
central densities, and thus higher temperatures, both of which act to
decrease $\lambda_{\nu}$,  (2) Hyperons decrease the  baryon degeneracy in the
central regions, and since the neutrinos couple quite  strongly to the
hyperons \cite{RP},  $\lambda_{\nu}$ is reduced; and (3) Hyperonization is 
accompanied by compressional heating and neutrino
production, both of which delay the cooling and deleptonization times.
Hyperons thus always act to increase the 
deleptonization and cooling times. Fig. 2 also shows  that a softer 
EOS favors longer diffusion times due to higher temperatures 
and densities in the inner regions of the star.\
To discriminate observationally between 
these different dense matter scenarios, the neutrino 
luminosities must be folded with the response of terrestrial detectors.
\vspace{-0.3in}
\begin{figure}[h]
\begin{center}
\epsfxsize=5.5in 
\epsfysize=3.in
\epsffile{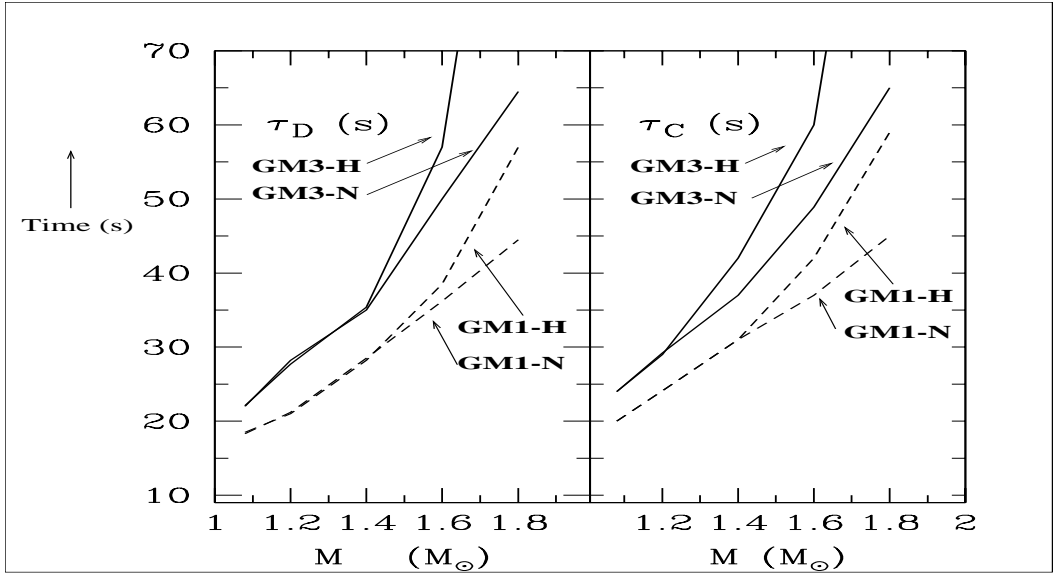}
\caption[]{\footnotesize Deleptonization and cooling times. Left panel:
$\tau_D$ for various models as a function of the initial baryon mass. Right
Panel: $\tau_C$ for the different models.} 
{\label{tau}}
\end{center}
\vspace{-0.2in}
\end{figure}


\vspace{-0.4in}
\begin{thebibliography}{99}
\itemsep=0cm
\bibitem{BL}
A.~ Burrows and  J.M.~Lattimer, ApJ.~{\bf 307} (1986) 178;
\bibitem{KJ}
W.~Keil and H.T.~Janka, Astron. \& Astrophys.~{\bf 296} (1995) 145;
\bibitem{RPL}S. Reddy, M. Prakash and J. M. Lattimer,
{Phys. Rev.} {\bf D} (1997) submitted.
\bibitem{RP}S. Reddy and M. Prakash,
{Astrophys. Jl.} {\bf 423} (1997) 689
\bibitem{S}R. F. Sawyer, Phys. Rev.~{\bf D11} (1975) 2740;
{Phys. Rev.} {\bf C40} (1989) 865.
\bibitem{IP}N. Iwamoto and C. J. Pethick, Phys. Rev.~{\bf D25} (1982) 313.
\bibitem{PR} M. Prakash, et. al, {Phys. Rep} {\bf 280} (1997) 1.
\bibitem{GM}N. K. Glendenning and S. A. Moszkowski, Phys. Rev. Lett.~{\bf 67} (1991) 2414.
\end{thebibliography}
\end{document}